\documentclass[twocolumn, aps, prb]{revtex4}
\usepackage{graphicx}
\usepackage{amssymb}
\usepackage{amsmath}
\usepackage{color}

\def\gsim{\lower -0.3ex \hbox{$>$} \kern -0.75em \lower 0.7ex
\hbox{$\sim$}}
\def\lsim{\lower -0.3ex \hbox{$<$} \kern -0.75em \lower 0.7ex
\hbox{$\sim$}}

\def\Vec#1{{\bf #1}}
\def\GVec#1{\mbox{\boldmath $#1$}}
\def\vare{\varepsilon}

\begin{document}
%%%%%%%%%%%%%%%%%%%%%%%%%%%%%%%%%%%%%%%%%%%%%%%%%%%%%%%%%%%%%%%%%%%%%%%%%%%%%%%
%
%%%%%%%%%%%%%%%%%%%%%%%%%%%%%%%%%%%%%%%%%%%%%%%%%%%%%%%%%%%%%%%%%%%%%%%%%%%%%%%
\title{
Optical absorption of 
twisted bilayer graphene
with interlayer potential asymmetry
%Optical absorption of twisted bilayer graphene
%with varying Fermi energy and interlayer bias
}
%\author{Pilkyung Moon and Mikito Koshino}
%\affiliation{
%School of Computational Sciences, Korea Institute for Advanced Study,
%Seoul, 130--722, Korea,\\
%Department of Physics, Tohoku University, 
%Sendai, 980--8578, Japan}
\author{Pilkyung Moon}
\email{pilkyung.moon@nyu.edu}
\affiliation{
New York University Shanghai,
Pudong, Shanghai, 200120, China}
\author{Young-Woo Son}
\email{hand@kias.re.kr}
\affiliation{
Korea Institute for Advanced Study,
Seoul, 130--722, Korea}
\author{Mikito Koshino}
\email{koshino@cmpt.phys.tohoku.ac.jp}
\affiliation{
Department of Physics, Tohoku University, 
Sendai, 980--8578, Japan}
\date{\today}

\begin{abstract}
We investigate the band structure
and the optical absorption spectrum
of twisted bilayer graphenes
with changing interlayer bias
and Fermi energy simultaneously.
We show that
the interlayer bias lifts
the degeneracy of the superlattice Dirac point,
while the amount of the Dirac point shift
is significantly suppressed in small rotation angles,
and even becomes opposite to the applied bias.
We calculate the optical absorption spectrum
in various asymmetric potentials and Fermi energies,
and associate the characteristic spectral features
with the band structure. 
The spectroscopic features
are highly sensitive to the interlayer bias and the Fermi energy,
and widely tunable by the external field effect.
%We discuss the interlayer bias dependence
%of the interband absorption spectrum
%as well as the Fermi energy dependence
%of the absorption steps
%and intraband absorption peaks.
\end{abstract}
\maketitle

%%%%%%%%%%%%%%%%%%%%%%%%%%%%%%%%%%%%%%%%%%%%%%%%%%%%%%%%%%%%%%%%%%%%%%%%%%%%%%%
%
%%%%%%%%%%%%%%%%%%%%%%%%%%%%%%%%%%%%%%%%%%%%%%%%%%%%%%%%%%%%%%%%%%%%%%%%%%%%%%%
\section{Introduction}

Twisted bilayer graphene (TBG) is
%a stack of two graphene layers
%coupled in an in-plane rotation angle
%a bilayer of graphene
a stacked and rotated two-layer graphene
with an in-plane rotation angle
other than
the integer multiples of $60^\circ$.
\cite{hass2007structural,hass2008multilayer,luican2011single}
With decreasing the rotation angle,
the misorientation between
two lattice periods produces
%The moir\'{e} interference
%between the misoriented lattices
%of the two layers of TBG
%gives rise to
a moir\'{e} interference pattern,
%with a long spatial period.
%superlattice potential
of which the spatial period
widely varies with
%rotation angles.
%the lattice registry.
the rotational alignment.
\cite{hermann2012periodic}
%Due to the moir\'{e} interference
%between two misoriented lattices,
%the interlayer registry
%varies over a long distance,
%producing a superlattice potential
%in the graphene.
%The beating of the mismatched lattices leads to the formation of a moire pattern with wavelength lambda(theta) that can be much larger than the lattice constant.
%Since such the additional periodicity
%is inherent to
%the interlayer registry of
%As the superlattice periodicity
%is introduced by
Due to the band-folding by the long-period potential,
TBG exhibits a peculiar band structure
with a renormalized Fermi velocity and a reduced saddle point energy,
which is distinctly different from monolayer graphene
and also from regularly-stacked  bilayer graphenes.
\cite{lopes2007graphene,hass2008multilayer,ni2008reduction,morell2010flat,shallcross2010electronic,trambly2010localization,bistritzer2011moirepnas,brihuega2012unraveling,sato2012zone,correa2014optical}
In addition,
the high-quality
superlattice potential,
which is inherent to
the lattice-mismatched stacking
of planar crystals,
can offer a unique opportunity
to investigate
%the competition
%between Bragg reflection
%and Landau quantization.
the self-similar energy spectrum
of charged particles
under the simultaneous influences of
a periodic potential and a magnetic field.
\cite{moon2012energy,moon2013opticalproperties,
bistritzer2011moireprb,wang2012fractal,
dean2013hofstadter,hunt2013massive,ponomarenko2013cloning}
%Recently
%it was theoretically proposed
%and experimentally confirmed
%In addition,
%it is shown that
%the electronic properties of TBG,
%such as a Fermi velocity and
%a saddle point energy,
%vary with the rotation angle
%between two layers.
%\cite{lopes2007graphene,trambly2010localization,shallcross2010electronic,morell2010flat,bistritzer2011moirepnas}

As a superlattice,
the optical absorption peak of TBG
systematically shift over
a wide range of wavelength with the rotation angle,
%regardless of the exact lattice commensurability.
suggesting that
%the spectroscopic characteristics
%serve as a fingerprint to
%identify the stacking angle.
this structure is a promising candidate
for optoelectronic applications.
\cite{moon2013opticalabsorption,wang2010stacking,stauber2013optical,
havener2013hyperspectral,havener2014van,
tabert2013optical,liang2014strongly}
%wang2010stacking,chen2011stacking,tabert2013optical
However, the effect of the interlayer bias
(i.e., the electrostatic potential difference between layers)
on the optical properties of TBGs
has not yet been investigated.
The interlayer bias has been widely used
in the band structure engineering
of multilayer graphene systems.
\cite{mccann2006asymmetry,castro2007biased,koshino2009gate,mak2009observation,zhang2009direct,castro2010electronic}
%In many cases, the Fermi level
%of graphene-related materials
%appears in the energy
%away from the charge neutrality point.
For TBG, it is reported that
the interlayer bias gives rise to some novel properties
in the band structure,
such as the additional renormalization of Fermi velocity
and topologically protected helical modes.
\cite{xian2011effects,san2013helical}
%Also, the optical
%\cite{wang2008gate,li2008dirac,liu2011graphene}
%and mechanical properties
%\cite{si2012electronic,woo2013ideal}
%of graphene can be controlled by
%shifting the Fermi energy
%from the charge neutrality point.

%In this paper,
%we investigate
%the effects of charge doping
%by varying Fermi energies.
%Due to the folding
%of Brillouin zone,
%the band structure
%and physical properties of TBG
%considerably deviate from
%the monotonous one of
%monolayer graphene,
%even at low carrier densities.

%While being a bilayer structure,
%the low-energy spectrum of TBG
%%resembles
%retains the linear band dispersion
%of monolayer graphene.
%\cite{lopes2007graphene,hass2008multilayer}
%However, 

The purpose of this work
is to reveal the band structure and the optical absorption spectrum of TBGs
under interlayer bias and charge doping.
%Here we investigate
%the band structure and optical absorption spectrum
%of TBGs in presence of the interlayer bias.
%we show that
%the absorption spectrum of TBG
%dramatically varies
%as Fermi energy
%approaches the saddle point:
%both the conductivity step
%and one of the absorption peaks
%of interband transition vanish
%while a strong absorption peak arises
%in an intra-conduction band transition.
The low-energy spectrum of TBG
is composed of four Dirac cones
originating from monolayer, 
and the Dirac point (electron-hole band touching point) 
are relatively shifted in energy
by applying an interlayer bias.
\cite{lopes2007graphene,xian2011effects,san2013helical}
%the monolayerlike spectrum
%in a low-energy regime
%of unbiased TBG
%is turned into
%the spectrum of
%non-degenerate Dirac cones
As a rotation angle reduces, however, we find that 
the Dirac point shift is strongly suppressed and
even becomes opposite to the case
when the interlayer coupling was absent.
In the angle below $2^\circ$,
the band structure is not simply regarded as
the combination of Dirac cones any more,
while the band touching at $K$ and $K'$ always remains
even in the interlayer bias, 
owing to the $C_2$ rotation symmetry.
%to the applied bias.
%We present an analytic
%expression for the amount of the shift,
%and show the condition
%to see the reverse of the shift.
We calculate the optical absorption spectra of TBGs
for various interlayer biases and Fermi energies,
and associate the characteristic absorption peaks and steps
with the specific properties in the band structure.
We find these spectroscopic features
strongly depend on the interlayer bias and the Fermi energy,
and thus widely tunable by the external gate electric field.
%In addition,
%we discuss the interband absorption step,
%which is linearly proportional to
%the interlayer bias,
%and the unique intraband absorption peaks
%that arise when Fermi energy
%deviates from charge neutrality point.
%We show that the optical spectrum of TBG with $\theta=1.47^\circ$
%are quite distinct from those of $\theta>2^\circ$
%due to the significant distortion of the band structure.

This paper organizes as follows.
Sec.\ II presents
our theoretical methods
utilizing a tight-binding Hamiltonian
on explicit lattice models of TBGs.
In Sec.\ III, we investigate
the band structures of TBGs
for various interlayer bias.
And in Sec.\ IV, we discuss
the characteristic optical
absorption spectrum of TBGs
while changing Fermi energy
as well as interlayer bias.
We conclude in Sec.\ V.

\section{Theoretical methods}

\subsection{Atomic structure}
\begin{figure*}
\begin{center}
\leavevmode\includegraphics[width=0.9\hsize]{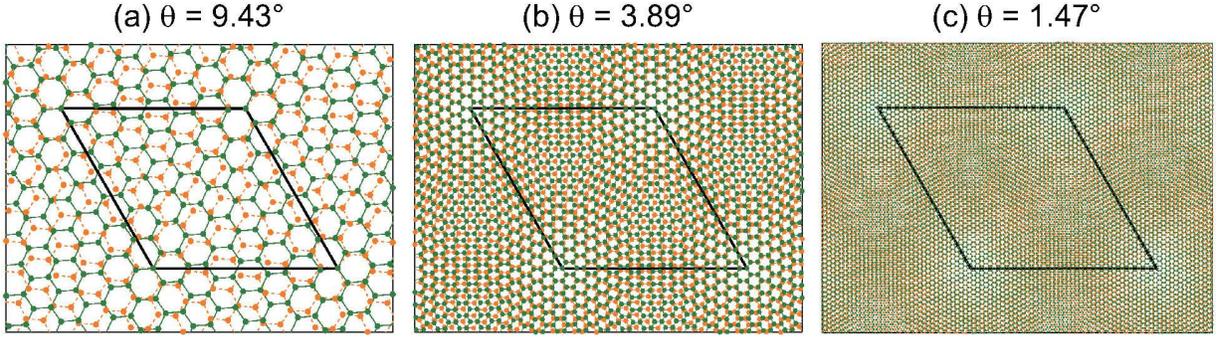}
\end{center}
\caption{(Color online) Atomic structures of TBGs with
(a) $\theta= 9.43^\circ$ and (b) $\theta=3.89^\circ$.
Dashed (orange) and solid (green) lines represent
the lattices of layers 1 and 2, respectively.
Black solid lines show the
moir\'{e} superlattice cell,
of which lattice constants
are $1.50\, \mathrm{nm}$,
$3.62\, \mathrm{nm}$, and
$9.59\, \mathrm{nm}$, respectively.
}
\label{fig_atomic_structures}
\end{figure*}

%In a moir\'{e} superlattice,
%we can define two different kinds of
%superlattice periods;
%(i) the size of the quasi-periodic
%moir\'{e} interference pattern
%%which continuously varies with
%%rotation angles,
%and
%(ii) the rigorous unit cell size of
%the exact coincidence lattice.
%%which discontinuously varies with the angles.
%The two periods exactly coincide
%only at specific discrete values of rotation angles,
%since the former continuously varies with
%rotation angles,
%while the latter does not.

In the TBG, the lattice structures
of two layers can be commensurate or incommensurate
sensitively depending on the rotation angle.
%However, the basic physical property is determined by the
%period of the moir\'{e} interference pattern
%regardless of the rigorous commensurability,
%and it almost continuously evolves
%with the rotation angle. 
%\cite{moon2013opticalabsorption}
However, the basic physical property
almost continuously
evolves with the period of the moir\'{e} interference pattern
which continuously varies
with the rotation angle.\cite{moon2013opticalabsorption}
Here we consider 
three specific commensurate TBG's $\theta = 9.43^\circ$,
$3.89^\circ$, and $1.47^\circ$, 
%representing short, middle, and long moir\'{e} periods.
which are illustrated in Fig.\ \ref{fig_atomic_structures}.
The dashed (orange) and
solid (green) lines
represent the lattices of layers 1 and 2.
Since the effects of
the relative displacement
%at the center of rotation
between two layers
on the band structure of TBG
are almost negligible,
\cite{moon2013opticalabsorption}
we only consider the structure
where the two layers share the atomic position
at the center of rotation.
%with $A-A$ coincidence at the center
%of rotation.

\subsection{Tight-binding model}
In studying the optical properties of TBG,
its lattice structures
should be included explicitly
into the model Hamiltonian
to catch the hidden symmetry
of the system correctly.
\cite{moon2013opticalabsorption}
To calculate the eigenenergies
and eigenfunctions of TBG,
we used the single-orbit ($p_z$)
tight-binding model,
where the hopping integral
$t(\Vec{R}_i - \Vec{R}_j)$
between any of two carbon atoms
at $\Vec{R}_i$ and $\Vec{R}_j$
are described by a conventional
Slater-Koster formula
with three parameters;
\begin{equation}
-t(\Vec{d}) = 
V_{pp\pi}(\Vec{d})\left[1-\left(\frac{\Vec{d}\cdot\Vec{e}_z}{d}\right)^2\right]
+ V_{pp\sigma}(\Vec{d})\left(\frac{\Vec{d}\cdot\Vec{e}_z}{d}\right)^2,
\label{eq_transfer_integral}
\end{equation}
where
$V_{pp\pi}(\Vec{d}) = V_{pp\pi}^0
\exp \left(- (d-a_0)/\delta_0\right)$,
$V_{pp\sigma}(\Vec{d}) = V_{pp\sigma}^0
\exp \left(- (d-d_0)/\delta_0\right)$,
$\Vec{d} = \Vec{R}_i - \Vec{R}_j$,
$d = |\Vec{d}|$,
the nearest intralayer coupling
$V_{pp\pi}^0 \approx -2.7\, \rm eV$,
the nearest interlayer coupling
$V_{pp\sigma}^0 \approx 0.48\, \rm eV$,
and the decay length
of the hopping integral
$\delta_0 \approx 0.184a$.
\cite{trambly2010localization,nakanishi2001conductance,
uryu2004electronic,slater1954simplified}
Here,
$a_0 = a/\sqrt{3} \approx 0.142$nm is the distance of
neighboring $A$ and $B$ sites on monolayer,
and $d_0 \approx 0.335\,\mathrm{nm}$
is the interlayer spacing.
For TBGs,
we consider the hopping
within $d < 4a_0$,
while for monolayer, $AA$, and $AB$,
we consider only
the nearest neighbor hopping.
The Hamiltonian is written as
\begin{eqnarray}
 H = -\sum_{\langle i,j\rangle}
t(\Vec{R}_i - \Vec{R}_j)
|\Vec{R}_i\rangle\langle\Vec{R}_j|
+ \sum_{i} V_i |\Vec{R}_i\rangle\langle\Vec{R}_i|
+ {\rm H.c.},
\label{eq_Hamiltonian_TBG}
\end{eqnarray}
where $|\Vec{R}_i\rangle$ 
represents the atomic state at site $i$,
and local on-site energy
to include an effect of layer-dependent
electric potentials in the presence of
transverse electric field.
For TBG with separate top
and bottom gate geometry,
\cite{nilsson2007transmission}
the system can be doped with
different electric potential
for each layer.
We model such a situation
with a rigid band approximation
while varying $V_i$.

\subsection{Dynamical conductivity}
We calculate the dynamical conductivity
\begin{eqnarray}
\sigma_{xx}(\omega) &=& 
\frac{e^2\hbar}{i S}
\sum_{\alpha,\beta}
\frac{f(\varepsilon_\alpha)-f(\varepsilon_\beta)}
{\varepsilon_\alpha-\varepsilon_\beta}
\frac{|\langle\alpha|v_x|\beta\rangle|^2}{\varepsilon_\alpha-\varepsilon_\beta+\hbar\omega+i\eta},
\label{eq_dynamical_conductivity}
\end{eqnarray}
from the eigenstate
$|\alpha\rangle$ ($|\beta\rangle$)
obtained by the tight-binding model.
Here, the sum is over all states,
$S$ is the area of the system, 
$f(\varepsilon)$ is the Fermi distribution function,
$\varepsilon_{\alpha}$ ($\varepsilon_{\beta}$)
represents the eigenenergy of the system,
$v_x=-(i/\hbar)[x,H]$ is the velocity operator, and 
$\eta$ is the phenomenological broadening
which is set to $0.13\,\rm meV$
in the following calculations.
A finite broadening factor $\eta$
is necessary to avoid a singular behavior
in the numerical calculation,
and here we set it to a sufficiently small value
to simulate the nearly ideal system.
A different choice of $\eta$
will change the broadening width
of the absorption spectra,
are not much sensitive.
Actual value of $\eta$ in the realistic situation
depends on the quality of the sample
and also on the experimental details.
The transmission of light
incident perpendicular
to a two-dimensional
system is
given by\cite{ando1975theory}
\begin{equation}
 T = \Big| 1+ \frac{2\pi}{c}\sigma_{xx}(\omega) \Big|^{-2}
\approx 1- \frac{4\pi}{c}{\rm Re}\,\sigma_{xx}(\omega).
\label{eq_Transmission}
\end{equation}

\section{Band structures}

\begin{figure*}
\begin{center}
\leavevmode\includegraphics[width=0.9\hsize]{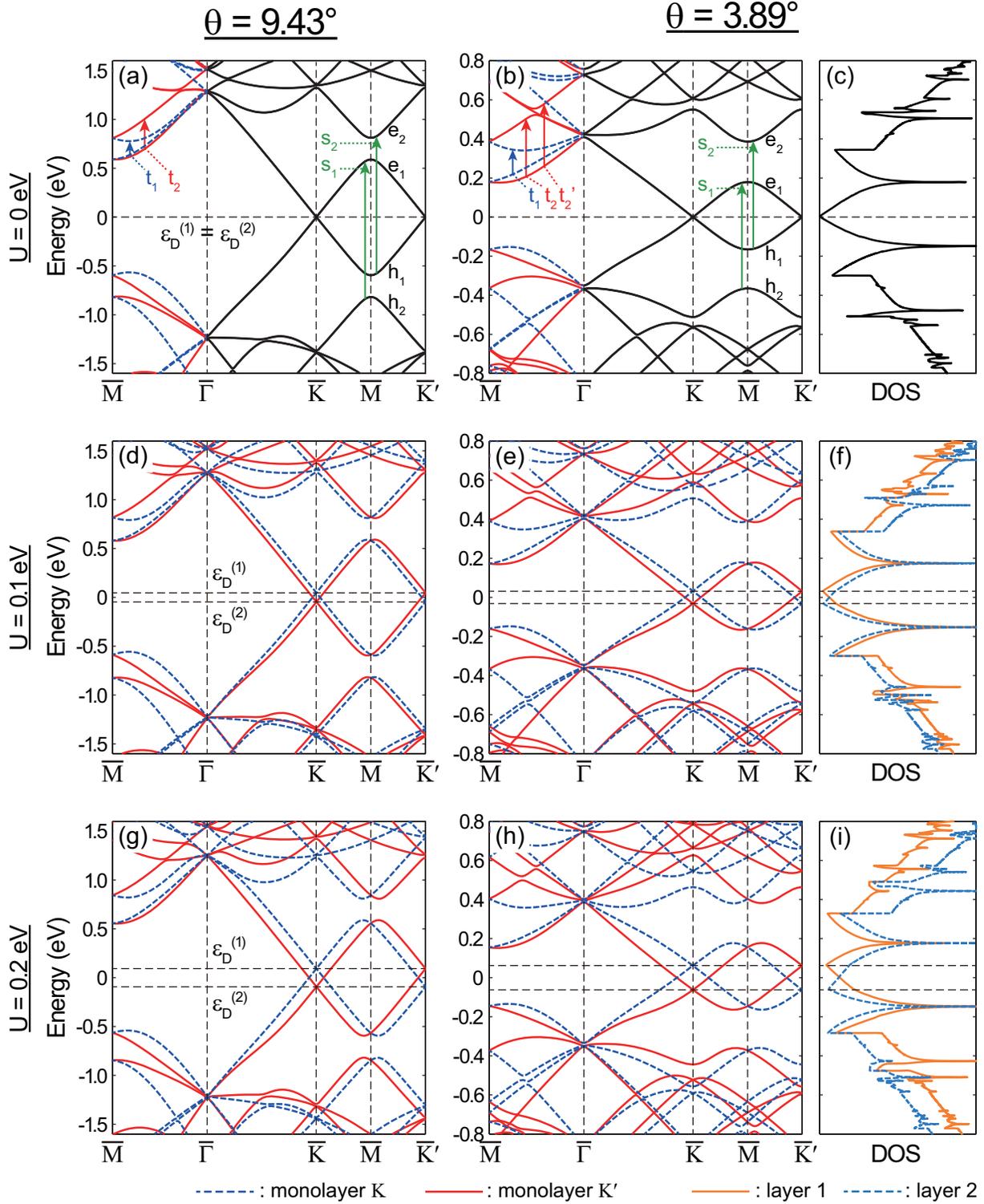}
\end{center}
\caption{(Color online) Band structures of
TBGs with $\theta = 9.43^\circ$
[(a), (d), and (g)]
and $\theta = 3.89^\circ$
[(b), (e), and (h)]
for different interlayer bias
$U=0\,\rm eV$,
$U=0.1\,\rm eV$, and
$U=0.2\,\rm eV$.
Dashed (blue) and solid (red) lines
represent the nearly degenerate branches
from the monolayer's
$K$ and $K'$ region, respectively
(see text).
(c), (f), and (i) show
the DOS of layer 1 (solid orange)
and 2 (dashed blue) in
$\theta = 3.89^\circ$
with $U$ = 0, 0.1, and 0.2\,\rm eV,
respectively.
$\vare_\mathrm{D}^{(1)}$ and $\vare_\mathrm{D}^{(2)}$
denote the Dirac point energies
of layer 1 and 2.
}
\label{fig_band_structures_TBGs}
\end{figure*}

\begin{figure}
\begin{center}
\leavevmode\includegraphics[width=0.8\hsize]{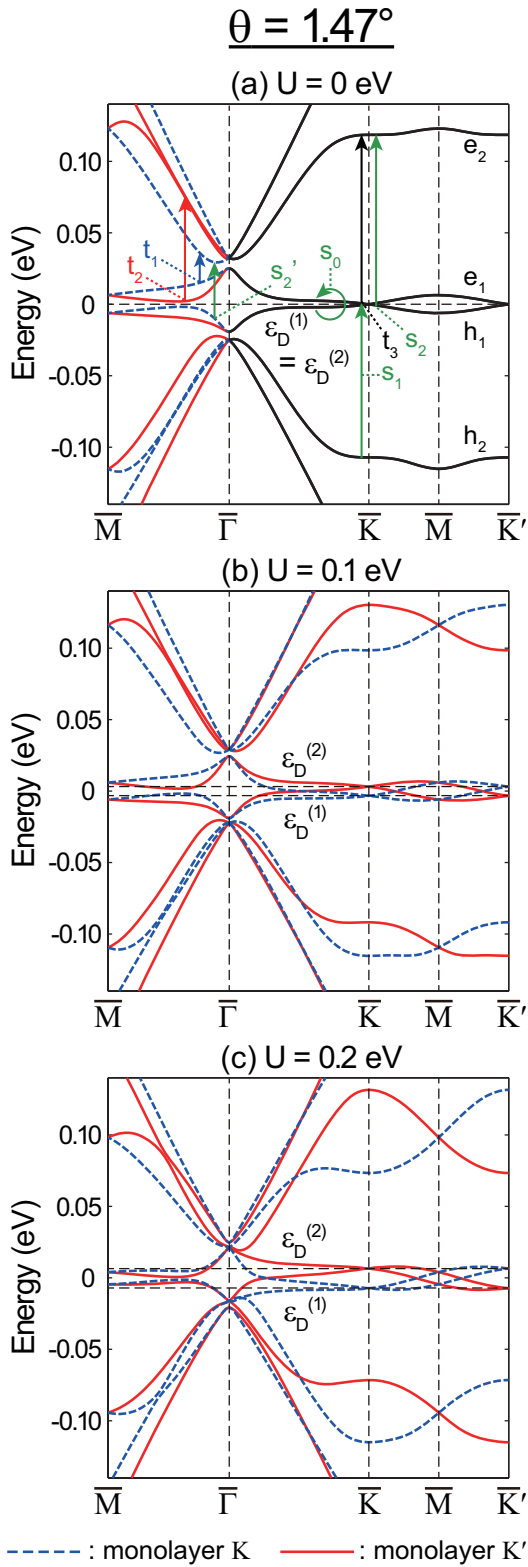}
\end{center}
\caption{(Color online)
Plots similar for Fig.\ \ref{fig_band_structures_TBGs}
for TBG with $\theta = 1.47^\circ$
with different interlayer bias
(a) $U=0\,\rm eV$,
(b) $U=0.1\,\rm eV$, and
(c) $U=0.2\,\rm eV$.
Note that the energy shifts
of the monolayer's $K$ and $K'$
branches with $U$
are opposite to those of
the structures in Fig.\ \ref{fig_band_structures_TBGs}
(see text).
}
\label{fig_band_structures_TBGs_1.47}
\end{figure}

\begin{figure}
\begin{center}
\leavevmode\includegraphics[width=0.9\hsize]{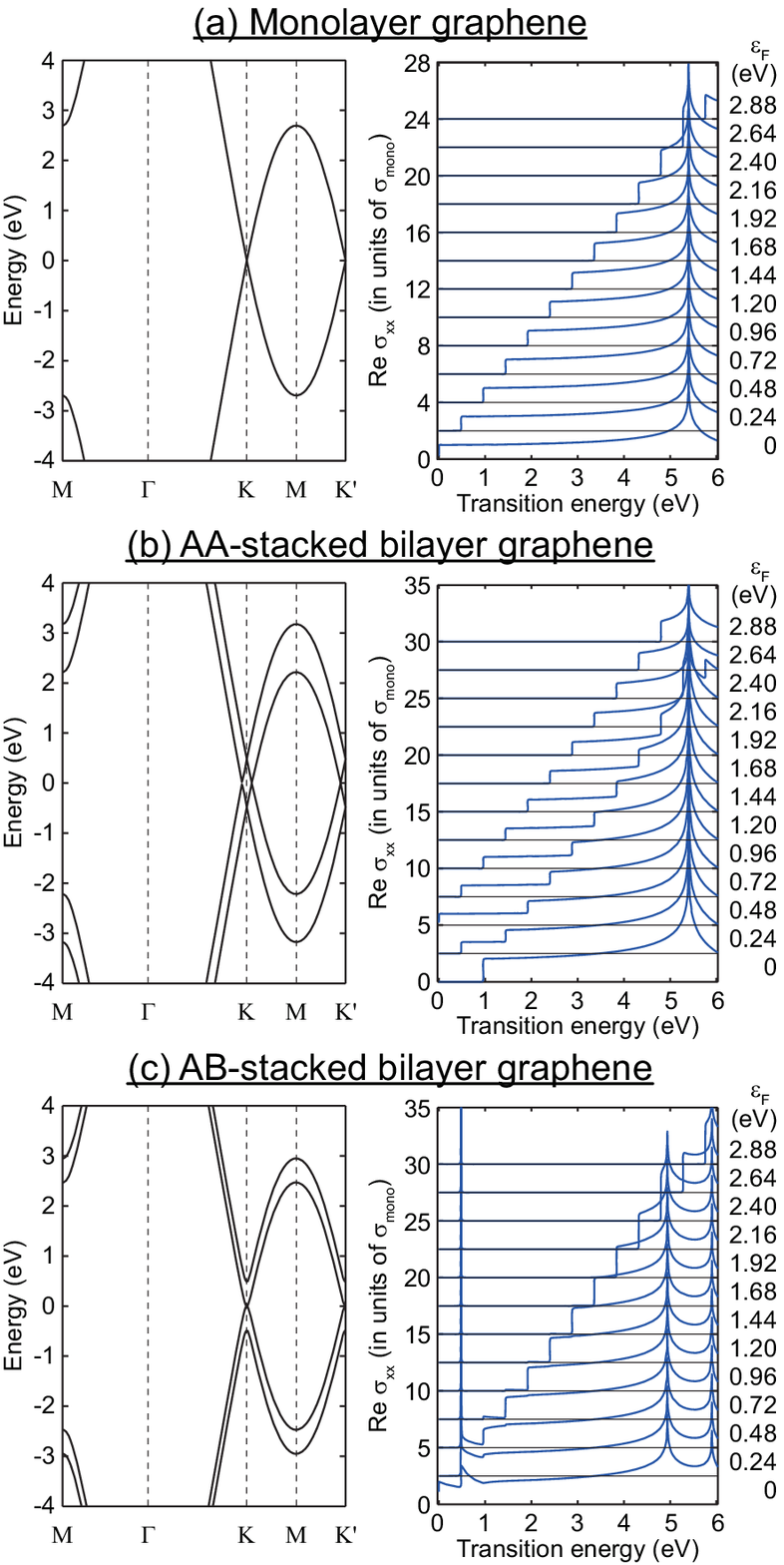}
\end{center}
\caption{
Band structure (left panel) and
dynamical conductivities
at various Fermi energies $\vare_\mathrm{F}$
(right panel)
for (a) monolayer graphene,
(b) $AA$-stacked bilayer graphene,
and (c) $AB$-stacked bilayer graphene.
}
\label{fig_regular_structures}
\end{figure}

% Band structures; symmetric

We calculate the band structure of TBG 
with the electrostatic potential
$U/2$ and $-U/2$ applied to the layer 1 and 2, respectively.
Figure \ref{fig_band_structures_TBGs}
shows the band structures
of TBGs with $\theta = 9.43^\circ$ and $3.89^\circ$
at $U=0,$ 0.1, $0.2\,\rm eV$,
and Fig.\ \ref{fig_band_structures_TBGs_1.47}
shows similar plots for $\theta=1.47^\circ$.
We also plot the density of states (DOS)
of $\theta = 3.89^\circ$
in Fig.\ \ref{fig_band_structures_TBGs}(c).
In all the plots, 
the charge neutrality point is set to $0\,\rm eV$.
Four valleys
$K^{(l)}$ and $K'^{(l)}$
of the layer $l$ $(l=1,2)$
are mapped to
the two superlattice Dirac points
$\bar{K}$ and $\bar{K}'$.
\cite{shallcross2010electronic}
In a low-energy regime,
a Dirac cone in one layer
strongly interacts with
only one Dirac cone in another layer,
since the other cones
are too far separated to be mixed
by the slowly varying potential
of the superlattice.
Thus, each energy band can be classified by
the monolayer's valley character,
i.e., $K$ band from $K^{(1)}$- and $K^{(2)}$-points,
and $K'$ band from $K'^{(1)}$- and $K'^{(2)}$-points,
\cite{moon2013opticalabsorption}
while it is folded in the common reduced Brillouin zone.
We marked the bands
with different colors
in Figs.\ \ref{fig_band_structures_TBGs}(a)
and \ref{fig_band_structures_TBGs}(b),
according to their original valley characters $K$ and $K'$.
These two bands are mirror symmetric to
each other with respect to the lines of
$\bar{K}-\bar{\Gamma}$,
$\bar{K}'-\bar{\Gamma}$,
and $\bar{K}-\bar{K}'$.
\cite{moon2012energy}
%reflecting the $C_2$ symmetry in the real-space lattice structure.
For comparison,
we plot the band dispersion of
monolayer graphene
and two forms of regular bilayer graphene,
$AA$ and $AB$,
at the left panels
in Figs.\ \ref{fig_regular_structures}(a),
\ref{fig_regular_structures}(b), and
\ref{fig_regular_structures}(c), respectively.
Compared to
the Dirac cones with offset in $AA$,
and also to the massive dispersion in $AB$,
TBG exhibits
monolayerlike dispersion
in the vicinity of
the charge neutrality point.
\cite{lopes2007graphene,hass2008multilayer}

The low-energy band structure of 
TBG is composed of the Dirac cones
originating from the two monolayer graphenes.
Due to the band folding, 
the band widths of the lowest conduction band ($e_1$)
and valence band ($h_1$)
of TBG gradually decrease as the rotation angle reduces.
\cite{moon2012energy}
As the angle is further reduced,
the energy scale of the folded band
becomes comparable to the interlayer coupling energy.
In $\theta = 1.47^\circ$ (Fig.\ \ref{fig_band_structures_TBGs_1.47}), 
the renormalized band velocity is significantly reduced
so that a flat band arises near the Dirac point.
\cite{lopes2007graphene,trambly2010localization,shallcross2010electronic,morell2010flat,bistritzer2011moirepnas}

% band structures with asymmetry

\begin{figure*}
\begin{center}
\leavevmode\includegraphics[width=0.9\hsize]{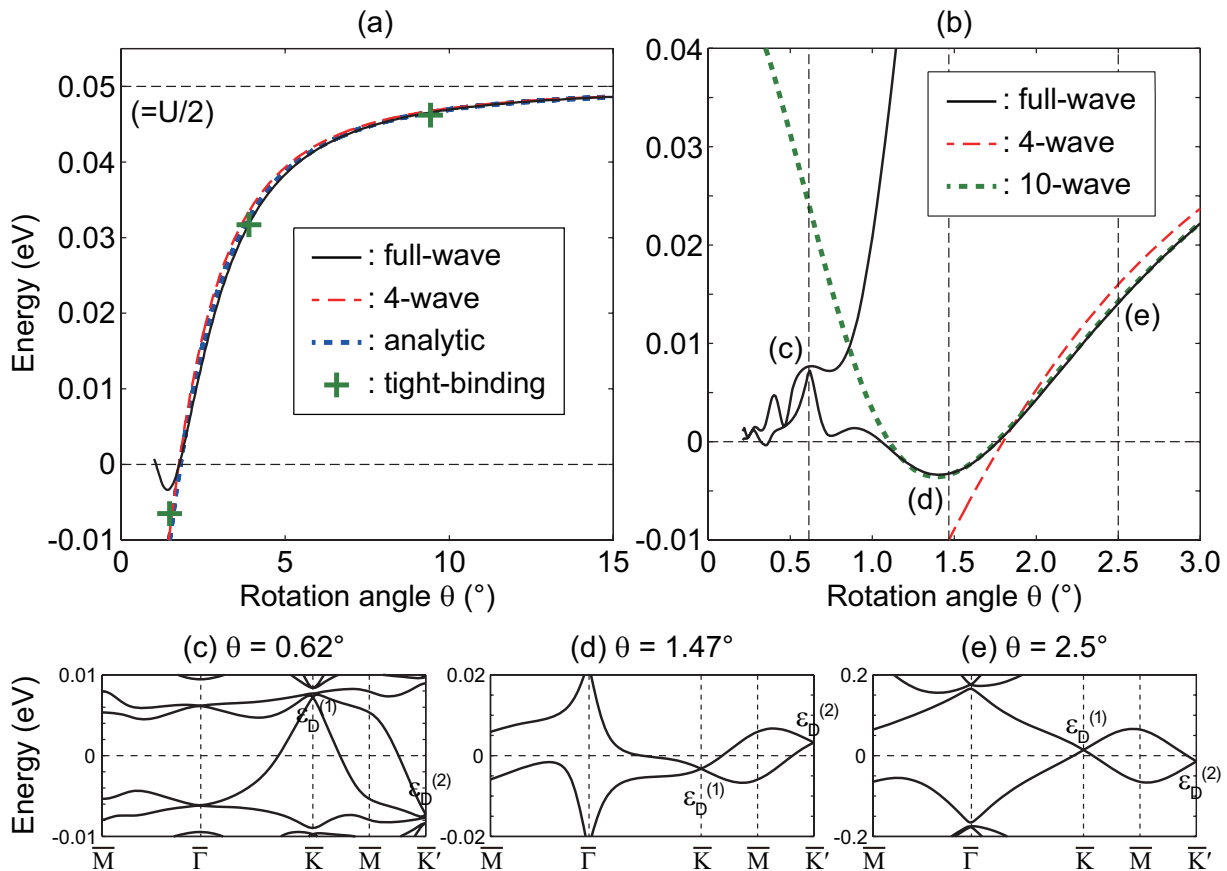}
\end{center}
\caption{
Dirac point energy $\vare_\mathrm{D}^{(1)}$
of TBG with interlayer bias $U=0.1\,\rm eV$,
as a function of the rotation angle
in (a) wide and (b) narrow ranges.
In (b), we also plot
the higher energy band
calculated by the full-wave model.
The bottom panels show the
band structures (showing only the bands from the monolayer's $K$ valley)
at $U=0.1\,\rm eV$ for (c) $\theta = 0.62^\circ$,
(d) $\theta = 1.47^\circ$, and
(e) $\theta = 2.5^\circ$,
calculated by the effective continuum model.
}
\label{fig_dirac_point_shift}
\end{figure*}

%Figures \ref{fig_band_structures_TBGs}(d),
%\ref{fig_band_structures_TBGs}(e),
%and \ref{fig_band_structures_TBGs_1.47}(b)
%[Figs.\ \ref{fig_band_structures_TBGs}(g),
%\ref{fig_band_structures_TBGs}(h),
%and \ref{fig_band_structures_TBGs_1.47}(c)]
%show the band structures of TBG
%with $\theta = 9.43^\circ$, $3.89^\circ$,
%and $1.47^\circ$, respectively,
%for $U = 0.1$ ($0.2$) $\rm eV$.

When the interlayer bias $U$ is introduced,
it lifts
the degeneracy of the superlattice Dirac point.
In large twist angles $\theta > 10^\circ$,
the energy shift of the Dirac point
approximates $\vare_\mathrm{D}^{(l)} \approx \pm U/2$ 
for layer $l = 1, 2$, respectively,
as if the two graphene layer were fully decoupled.
\cite{lopes2007graphene,xian2011effects,san2013helical}
This is because in large $\theta$, 
the interlayer coupling is so weak near zero energy that
the Dirac cones of two layers remain almost independent.
As the rotation angle decreases,
however, the energy offset between
the two Dirac points is suppressed.
For $U = 0.1$ and $0.2\,\rm eV$,
TBG with $\theta = 9.43^\circ$ shows
$\vare_\mathrm{D}^{(1)} = 46.2$ and $92.3\,\rm meV$
($\approx 8\%$ suppression),
while TBG with $\theta = 3.89^\circ$
shows $\vare_\mathrm{D}^{(1)} = 31.7$ and $61.5\,\rm meV$
($\approx 38\%$ suppression),
respectively.
In small $\theta$, the energy scale of the folded band
becomes comparable to $|U|$
so that the interlayer band mixing becomes prominent
near Dirac points.
Thus, the wave function is distributed
to both layers,
and even at the Dirac point,
the layer polarization
of the carrier is weakened
and the Dirac point shift is decreased.
As $\theta$ further reduces,
the shift of Dirac point
changes its sign,
as can be seen from
the negative $\varepsilon_D^{(1)}$ in
Fig.\ \ref{fig_band_structures_TBGs_1.47}(b)-(c).
%due to the significant mixing
%between the states in the two layers.

Using the effective continuum model,\cite{moon2013opticalabsorption}
the amount of the shift can be approximately estimated as
\begin{equation}
\varepsilon^{(1)}_{\rm D} = 
\pm\frac{U}{2} \frac{(\hbar v q)^2 - 6u_0^2}{(\hbar v q)^2 + 6u_0^2},
\label{eq_analytic}
\end{equation}
with $\pm$ for layer $l=1,2$, respectively,
where $u_0 \approx 0.11$eV is the interlayer coupling energy,
\cite{moon2013opticalabsorption} and 
$q = 8 \pi \sin(\theta/2)/(3a)$ 
is the length of the moir\'{e} reciprocal lattice vector.
This is obtained by
a few-mode approximation
where only $k$-points that directly couple to the Dirac point of layer 1
are taken into account
in the effective continuum model. \cite{moon2013opticalabsorption}
The detail of the derivation is presented in Appendix.
Equation (\ref{eq_analytic}) stands at large angle $\theta$ such that
$\hbar v q \gg u_0$, and also the moderate bias $|U| \ll u_0$.
%Equation (\ref{eq_analytic}) shows that
%the amount of the suppression
%($1-E/U$)
%is less sensitive to the applied bias $U$.
The analytic expression shows that
the Dirac point shifts monotonically reduces
as decreasing $\theta$ (i.e., decreasing $q$),
and reaches zero at $\hbar v q = \sqrt{6} u_0$,
or
\begin{equation}
\theta_c \approx  \frac{3 \sqrt{6} a}{4 \pi \hbar v} u_0
\approx 1.7^\circ.
\end{equation}
In Fig.\ \ref{fig_dirac_point_shift}(a),
we plot $\vare_\mathrm{D}^{(1)}$
for $U = 0.1\, \rm eV$
calculated by
the full-wave effective continuum model,
four-wave model [Eq.\ (\ref{eq_4wave})],
analytic model [Eq.\ (\ref{eq_analytic})],
and tight-binding method.
The four models are
almost perfectly consistent with each other
at $\theta > 2^\circ$.
%As $\theta$ decreases,
%the amount of the Dirac point shift reduces
%and even drops below zero.
We plot $\vare_\mathrm{D}^{(1)}$
at small angle regime
in Fig.\ \ref{fig_dirac_point_shift}(b).
Figures \ref{fig_dirac_point_shift}(c), (d), and (e)
show the band structures
from the monolayer's $K$ valley
at $\theta = 0.62^\circ, 1.47^\circ$, and $2.5^\circ$,
respectively,
calculated by the effective continuum model.
Equation (\ref{eq_analytic}) is no longer valid
for these cases
because the condition $\hbar v q \gg u_0$ does not hold.
In decreasing $\theta$, 
the Dirac point drops below zero
at $\theta \approx 1.7^\circ$,
but again increases and goes to the positive region
at $\theta \approx 1^\circ$.
These features can be reproduced by
considering additional six-waves 
to the four-wave model.
As $\theta$ further reduces from $\theta =1^\circ$,
the higher energy bands
begin to be mixed 
to the lowest band, and the energy band is no longer
described by the single Dirac cone [Fig.\ \ref{fig_dirac_point_shift}(c)].
There the Dirac point energy 
exhibits a complex oscillatory behavior.

% Dirac point protection

It is somewhat surprising to see that the Dirac points are never gapped
even in the presence of the interlayer coupling and the 
interlayer asymmetric potential.
Generally, it is known that 
the coexistence of the time reversal symmetry 
and the spatial inversion symmetry requires vanishing of
the Berry curvature at any non-degenerate points in the energy band, 
\cite{haldane2004berry, fu2007topological}
and this guarantees the robustness of band touching points
in two-dimensional systems. \cite{koshino2013electronic}
In a similar manner, we can show that
the coexistence of the time reversal symmetry
and the in-plane $C_2$ (180$^\circ$) rotation symmetry 
(instead of the inversion symmetry)
also concludes the same Dirac point protection in two dimensions, 
because the degree of freedom in $z$-direction
does not change the argument.
The lattice structure of TBG lacks the inversion symmetry 
but possesses the $C_2$ symmetry,\cite{moon2012energy}
and moreover, $C_2$ symmetry holds
even in the presence of the interlayer potential asymmetry
because $C_2$ does not flip the layers.
This is the origin of the Dirac point protection in the
asymmetric TBG.
In contrast, $AB$-stacked bilayer graphene has the inversion symmetry
but lacks the $C_2$ symmetry,
and thus the band touching 
is lifted by the interlayer potential difference. \cite{mccann2006landau}

%DOS

The interlayer band mixing also
influences the layer-wise DOS of the biased TBG.
In Figs.\ \ref{fig_band_structures_TBGs}(f)
and \ref{fig_band_structures_TBGs}(i),
we plot
the DOS for layer 1 and 2
by solid (orange) and dashed (blue) lines, respectively.
%The figure shows that
%the DOS for each layer of TBG
%exhibits a minimum
%at the Dirac point energy of each layer,
%as the same to
%that of monolayer graphene.
In a fully decoupled bilayer,
the DOS of the layer $l$
should vanish
at $\vare_\mathrm{D}^{(l)}$.
Due to the interlayer interaction,
however,
the layer-wise DOS of the biased TBG
does not completely vanishes
at the Dirac point of each layer.
The minimum of the layer-wise
DOS at each layer's Dirac point
becomes substantial
as the rotation angle reduces
or $|U|$ increases.

\section{Optical spectrum}

\begin{figure*}
\begin{center}
\leavevmode\includegraphics[width=0.9\hsize]{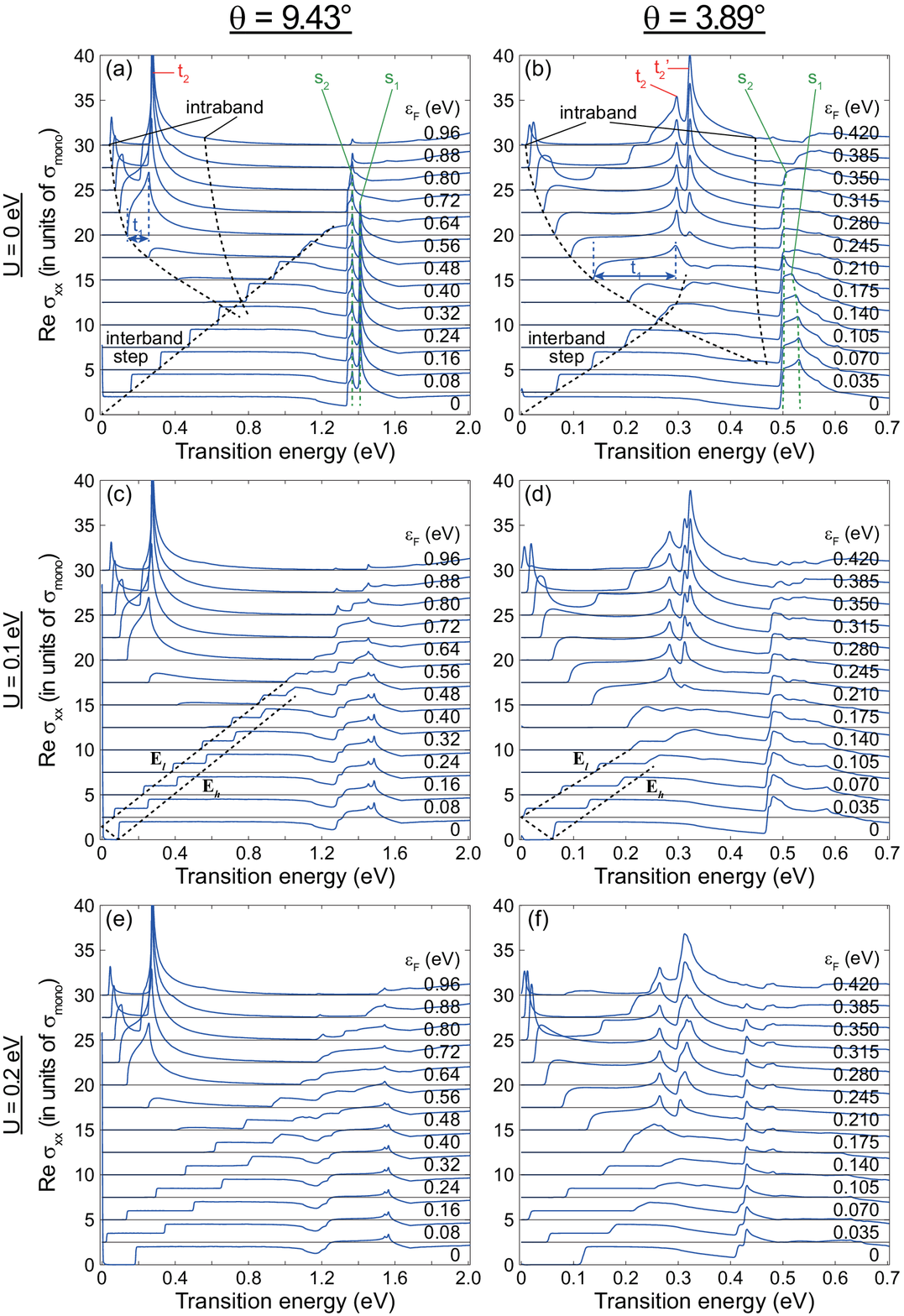}
\end{center}
\caption{
Dynamical conductivities of TBGs
with $\theta = 9.43^\circ$
with
(a) $U=0\,\rm eV$,
(c) $U=0.1\,\rm eV$,
(e) $U=0.2\,\rm eV$,
and $\theta = 3.89^\circ$
with
(b) $U=0\,\rm eV$,
(d) $U=0.1\,\rm eV$,
(f) $U=0.2\,\rm eV$,
for various Fermi energies $\vare_\mathrm{F}$.
}
\label{fig_optical_conductivities_TBGs}
\end{figure*}

\begin{figure}
\begin{center}
\leavevmode\includegraphics[width=0.9\hsize]{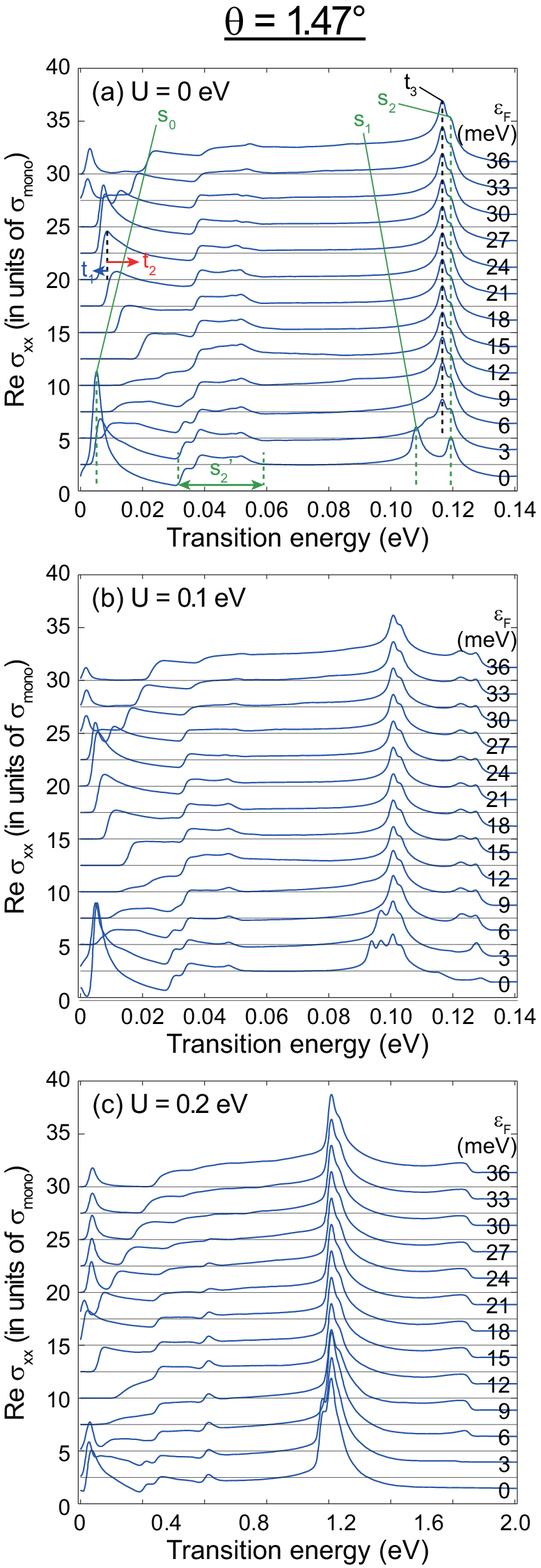}
\end{center}
\caption{
Plots similar for Fig.\ \ref{fig_optical_conductivities_TBGs}
for TBG with $\theta = 1.47^\circ$
with different interlayer bias
(a) $U=0\,\rm eV$,
(b) $U=0.1\,\rm eV$, and
(c) $U=0.2\,\rm eV$
for various Fermi energies $\vare_\mathrm{F}$.
}
\label{fig_optical_conductivities_TBGs_1.47}
\end{figure}

We calculated the optical conductivities of
the TBGs
for various Fermi energies $\varepsilon_F$ and interlayer asymmetry $U$.
We plot the optical absorption spectra
of TBGs with $\theta = 9.43^\circ$
for different interlayer bias of
$U = 0$, $0.1$, and $0.2\,\rm eV$
in Figs.\ \ref{fig_optical_conductivities_TBGs}(a),
\ref{fig_optical_conductivities_TBGs}(c),
and \ref{fig_optical_conductivities_TBGs}(e), respectively.
Similar figures for $\theta = 3.89^\circ$
are presented in Figs.\
\ref{fig_optical_conductivities_TBGs}(b),
\ref{fig_optical_conductivities_TBGs}(d),
and \ref{fig_optical_conductivities_TBGs}(f).
For comparison,
we plot the absorption spectra of
monolayer graphene and
regular bilayer graphenes ($AA$-stack and $AB$-stack)
at the right panels
in Figs.\ \ref{fig_regular_structures}(a),
\ref{fig_regular_structures}(b), and
\ref{fig_regular_structures}(c), respectively.
When Fermi energy lies at the charge neutrality point
($\vare_\mathrm{F} = 0$),
the low-energy optical spectrum of monolayer graphene
exhibits universal dynamical conductivity
\begin{eqnarray}
\sigma_{\rm mono} = \frac{g_v g_s}{16} \frac{e^2}{\hbar},
\end{eqnarray}
due to the linear dispersion of the band.
\cite{nair2008fine,ando2002dynamical,gusynin2006transport,
gusynin2006unusual}
Here $g_s = 2$ and $g_v = 2$
are the spin and valley ($K$, $K'$)
degeneracy, respectively.
Meanwhile,
$AA$-stacked bilayer at $\vare_F=0$
shows interband absorption step at $2V^0_{pp\sigma}$,
since Dirac points are located away from charge neutrality point.
\cite{xu2010infrared,tabert2012dynamical}
$AB$-stacked bilayer at $\vare_F=0$ exhibits
absorption edge at $V^0_{pp\sigma}$
which reflects the interband transition
from the low-energy band to the split band.
\cite{PhysRevB.75.155430,koshino2009electronic,
zhang2008determination,kuzmenko2009infrared,li2009band}

The optical spectrum of TBG
is characterized by several unique features described below.

(i) {\it Interband transition peaks associated with 
saddle points.}
In each panel, we see a characteristic peak
near $1.4\, \mathrm{eV}$ at $\theta = 9.43^\circ$
and $0.5\, \mathrm{eV}$ at $\theta = 3.89^\circ$.
They arise from the interband transitions
between the saddle point of
the lowest band
($h_1$ and $e_1$)
and the band edge
of the second band
($h_2$ and $e_2$), 
which are marked as $s_1$ ($h_2 \rightarrow e_1$)
and $s_2$ ($h_1 \rightarrow e_2$)
in Figs.\ \ref{fig_band_structures_TBGs}(a)
and \ref{fig_band_structures_TBGs}(b).
\cite{moon2013opticalabsorption}
%in both Figs.\ \ref{fig_band_structures_TBGs}(a)
%and \ref{fig_band_structures_TBGs}(b).
Here note that the direct transition between the saddle points
($h_1 \rightarrow e_1$) is forbidden by the selection rule.
\cite{moon2013opticalabsorption}
When $\vare_\mathrm{F}$ is increased to the saddle point,
one of the two interband peaks ($s_1$)
disappears
%[right arrows in Figs.\ \ref{fig_optical_conductivities_TBGs}(a)
%and \ref{fig_optical_conductivities_TBGs}(b)]
%while the other one ($h_1 \rightarrow e_2$) remains,
since the transition to
the occupied band $e_1$
is no longer possible.
In increasing
$U$, on the other hand,
we see the interband transition peaks
become broad and the intensity
gets significantly reduced.
This is because, in the presence of $U$, 
the saddle point $h_1$ ($e_1$) and the band edge $e_2$ ($h_2$)
shift in the opposite directions
as we can see from Fig.\ \ref{fig_band_structures_TBGs},
and this strongly affects the joint density of states
associated with the transition $s_1$ and $s_2$.

%and a considerable reduction of conductivity
%right below the peak energy.\cite{moon2013opticalabsorption}
%
%The peaks have highly asymmetric line shapes,
%since the DOS of the two bands
%that participate in the optical transition
%exhibit a sharp peak
%[$h_1$ and $e_1$ in Fig.\ \ref{fig_band_structures_TBGs}(c)]
%and a step
%[$h_2$ and $e_2$ in Fig.\ \ref{fig_band_structures_TBGs}(c)], respectively.
%The peak energies 
%sensitively depend on rotation angles,
%and due to the electron-hole asymmetry,
%the energies of the two peaks
%do not always coincide.
%Together with the reduction of DOS
%beyond the saddle points,
%the selection rule is responsible for
%the conductivity drop
%at the right below the peak energy.

(ii) {\it Interband absorption step.}
In the symmetric TBG $(U=0)$ at charge neutral ($\vare_F = 0$),
the optical conductivity is close to $2\sigma_{\rm mono}$
in the low frequencies.
As the Fermi energy $\vare_\mathrm{F}$ deviates from 0, however,
we have a discrete step
below which the absorption is absent, \cite{tabert2013optical}
because the filled electrons forbid the corresponding excitations.
The feature is analogous to monolayer graphene
[Fig.\ \ref{fig_regular_structures}(a)],
as it reflects the charging of Dirac cone.
The step linearly shifts to higher energies
until it vanishes when $\vare_\mathrm{F}$ reaches
the saddle point, where the linear dispersion is lost.

%In contrast to the spectrum of unbiased TBG
%and monolayer graphene,
%there appears an interband step
%in the low-energy spectrum
%of biased TBG
%%even when the Fermi energy
%%lies at the charge neutrality point.
%even at the charge neutrality condition.
%As the Fermi energy $\vare_\mathrm{F}$
%deviates from the charge neutrality point,

In the presence of the asymmetric potential $U$,
the interband absorption step splits into two different energies
\begin{eqnarray}
E_l = 2|\vare_\mathrm{F} - \vare_\mathrm{D}^{(1)}|,
\nonumber\\
E_h = 2|\vare_\mathrm{F} - \vare_\mathrm{D}^{(2)}|,
\end{eqnarray}
due to the relative shift of Dirac point energies 
$\varepsilon^{(1)}_{\rm D}$ and $\varepsilon^{(2)}_{\rm D}$.
The step positions $E_l$ and $E_h$ linearly depend on $U$
as expected from Eq.\ (\ref{eq_analytic}).
We see similar absorption steps
in $AA$-stacked bilayer [Fig.\ \ref{fig_regular_structures}(b)], 
\cite{tabert2012dynamical}
where the Dirac points are originally split at $U=0$
by the interlayer coupling energy $\sim V^0_{pp\sigma}$.
These steps can be moved by the interlayer asymmetry $U$,
while the shift is just proportional to $U^2$ unlike in TBG.

(iii) {\it Intraband absorption peaks.}
{
In the absorption spectrum
of charge-doped TBGs ($\varepsilon_F\neq 0$),
we see a series of peaks centered at $\sim 0.3\, \mathrm{eV}$,
which are clearly distinct from those arising from
saddle point interband transitions [(i)].
Those peaks reflect the intraband transition
$e_1 \rightarrow e_2$,
%which becomes available by
%the folding of Brillouin zone
%in large superlattices,
which is indicated more specifically as $t_1$ and $t_2$
in Figs.\ \ref{fig_band_structures_TBGs}(a)
and \ref{fig_band_structures_TBGs}(b).
We see that the two energy bands associated with $t_2$  
are almost parallel between $\bar{M}$ and $\bar{\Gamma}$,
where the transition energy is nearly equal to $0.3\, \mathrm{eV}$.
This actually causes a sharp absorption peak near $0.3\, \mathrm{eV}$.
On the contrary, the transition energy for $t_1$
significantly depends on
wavevector $k$, giving a broad absorption peak.
In $\theta = 3.89^\circ$, the $t_2$ intraband peak splits into 
the two peaks at $0.30\,\rm eV$ and $0.32\,\rm eV$,
reflecting the band anticrossing
near the mid-point between $\bar{M}$ and $\bar{\Gamma}$.
When the Fermi energy is increased above the saddle point
and the second conduction band becomes partially filled,
a region with no-absorption
appears in the broad $t_1$ peak,
%[left arrows in Figs.\ \ref{fig_optical_conductivities_TBGs}(a)
%and \ref{fig_optical_conductivities_TBGs}(b)],
since the transition to the occupied band $e_2$
becomes forbidden.
In increasing the interlayer asymmetry $U$,
we see that the structure of the intraband transition peak
is not considerably changed, 
in contrast to the significant broadening 
of the interband transition peaks (i).
This is consistent with the fact that
the energy dispersion associated the transitions $t_1$ and $t_2$
are not strongly modified by $U$
as we see in Fig.\ \ref{fig_band_structures_TBGs}.

The optical spectrum of TBG with $\theta = 1.47^\circ$
is quite different from those of $\theta > 2^\circ$
due to the significant distortion of the band structure.
We plot the optical absorption spectrum
of TBG with $\theta = 1.47^\circ$
for different interlayer bias of
$U = 0$, $0.1$, and $0.2\,\rm eV$
in Figs.\
\ref{fig_optical_conductivities_TBGs_1.47}(a),
\ref{fig_optical_conductivities_TBGs_1.47}(b),
and \ref{fig_optical_conductivities_TBGs_1.47}(c), respectively.
%s_0
The low-energy spectrum
of TBG with $\theta = 1.47^\circ$
at $\varepsilon_F = 0$
is characterized by
a unique absorption peak
near $5.3\, \mathrm{meV}$,
which is indicated as $s_0$
in Fig.\ \ref{fig_band_structures_TBGs_1.47}(a).
It is the transition
between the flat bands
that gives a strong absorption peak.
The peak suddenly disappears
as $\varepsilon_F$ deviates
from charge neutrality point
since the band width of $e_1$ is very narrow.
%s_1,s_2,t_3
We can also see two characteristic peaks
at high-energy spectrum,
which arise from the transitions
$s_1$ and $s_2$ in Fig.\
\ref{fig_band_structures_TBGs_1.47}(a).
In TBGs with $\theta=9.43^\circ$ and $3.89^\circ$,
the interband transition peaks (i)
occur at $\bar{M}$
where the band edge of the second band
($h_2$ and $e_2$)
resides there
[Figs.\ \ref{fig_band_structures_TBGs}(a)
and \ref{fig_band_structures_TBGs}(b)].
In TBG with $\theta=1.47^\circ$,
however,
the peaks occur near $\bar{K}$ and $\bar{K'}$,
at which the saddle points of the second band reside
[Fig.\ \ref{fig_band_structures_TBGs_1.47}(a)].
Likewise with the peak from $s_0$,
$s_1$ also vanishes
as $\varepsilon_F$ deviates
from charge neutrality point.
However, at that $\varepsilon_F$,
a new peak from the intraband transition
$t_3$ in Figs.\ \ref{fig_band_structures_TBGs_1.47}(a)
and \ref{fig_optical_conductivities_TBGs_1.47}(a)
is activated
at the energy close to $s_2$.
In increasing the interlayer asymmetry $U$,
each peak splits into two
by the degeneracy lift of the bands
(Fig.\ \ref{fig_band_structures_TBGs_1.47}).
In increasing $U$,
the lower-energy peak $s_1$
gradually redshift
and their intensities are
significantly enhanced.
%s_2'
The spectrum at around $0.45\, \mathrm{eV}$
originates from the transition
near $\bar{\Gamma}$
[$s_2'$ in Fig.\ \ref{fig_band_structures_TBGs_1.47}(a)].
The corresponding transitions
in TBGs with $\theta=9.43^\circ$ and $3.89^\circ$
have the energy higher than $s_1$ and $s_2$,
but in $\theta=1.47^\circ$,
it has much smaller energy than $s_1$ and $s_2$
due to the band folding.
%t_1,t_2
In the low-energy spectrum 
of charge-doped TBGs ($\varepsilon_F \ne 0$),
we see a series of peaks
that come from
the intraband transitions
$t_1$ and $t_2$ in Fig.\ \ref{fig_band_structures_TBGs_1.47}(a).
%\textcolor{blue}{@@@Following argument is not clear:}
%Here, $t_1$ gives a sharp peak
%by the transition to the band edge of
%the second band ($e_2$),
%while $t_2$ gives a broad absorption peak.

\section{Conclusion}
We investigated the band structure and 
the optical absorption spectrum of TBGs with varying
interlayer bias and Fermi energies theoretically.
We showed that the interlayer bias lifts
the degeneracy of the superlattice Dirac point,
while the shift of the Dirac point is significantly suppressed
as the interlayer rotation angle $\theta$ reduces.
The low-energy band structure including the Dirac point shift 
was analytically described by the effective continuum model.
We calculated the optical absorption spectrum
and associate the characteristic spectral features
with the band structure. The spectrum consists of
the interband and intraband transition peaks
as well as the interband absorption steps,
where the peak (step) positions and amplitudes
are highly sensitive to the interlayer bias and the Fermi energy.
Meanwhile, we showed that both the band structure
and optical spectrum of TBG with $\theta=1.47^\circ$
quite different from those of $\theta>2^\circ$,
due to the strong band distortion
caused by interlayer coupling at the low-energy regime.
Our calculation results as well as analysis
can clarify optical spectrum of TBGs in actual experimental setups
such as TBGs on top of different layered materials
or the systems in field effect transistor geometries.

\section*{ACKNOWLEDGEMENTS}
P.\ M.\ was supported by
New York University Shanghai (research funds)
and East China Normal University (research facilities).
Y.-W.\ S.\ was supported by the NRF of Korea
grant funded by the MSIP (CASE, 2011-0031640
and QMMRC, NO.\ R11-2008-053-01002-0).
M.\ K.\ is funded by JSPS Grant-in-Aid for Scientific Research No.
24740193, No. 25107005.
Computations were supported by the CAC of KIAS.

\appendix

\section{Derivation of the Dirac point shift}

Here we derive an approximate analytic expression
Eq.\ (\ref{eq_analytic})
for the energy shift of the Dirac points in the 
TBG under a asymmetric potential.
We adopt the effective continuum model, \cite{moon2013opticalabsorption}
and construct the Hamiltonian
only taking account of the $k$-points
that directly couple to the Dirac point of layer 1,
$\Vec{K}_\xi^{(1)}$
($\xi = \pm 1$ for monolayer's $K$ and $K'$, respectively).
In the effective model,
the monolayer state of the layer 1
at $\Vec{k}$ couples to
the state of the layer 2
at $\Vec{k}$,
$\Vec{k} - \xi \Vec{G}_1^{\rm M} - \xi \Vec{G}_2^{\rm M}$, and
$\Vec{k} - \xi \Vec{G}_2^{\rm M}$,
where $\Vec{G}_1^{\rm M}$ and $\Vec{G}_2^{\rm M}$
are moir\'{e} reciprocal vectors,
by the Fourier component
of the moir\'{e} superlattice potential.
The reduced Hamiltonian
of the four-wave approximation becomes
\begin{widetext}
\begin{eqnarray}
 H_{\rm eff}(\Vec{k}) = 
\begin{pmatrix}
H_\xi^{(1)}(\Vec{k})+U/2 & U_1^\dagger & U_2^\dagger & U_3^\dagger \\
U_1 & H_\xi^{(2)}(\Vec{k})-U/2 & 0 & 0 \\
U_2 & 0 & H_\xi^{(2)}(\Vec{k}-\xi \Vec{G}_1^{\rm M} - \xi \Vec{G}_2^{\rm M})-U/2 & 0 \\
U_3 & 0 & 0 & H_\xi^{(2)}(\Vec{k} - \xi \Vec{G}_2^{\rm M})-U/2
\end{pmatrix},
\label{eq_4wave}
\end{eqnarray}
\end{widetext}
with
\begin{eqnarray}
&& H_\xi^{(l)}(\Vec{k}) = -\hbar v (\Vec{k}-\Vec{K}_\xi^{(l)})\cdot \GVec{\sigma},
\nonumber\\
&& U_1 = u_0
\begin{pmatrix}
1 & 1 \\
1 & 1
\end{pmatrix},
\nonumber\\
&& U_2 = u_0
\begin{pmatrix}
1 & \omega^{-\xi} \\
\omega^{\xi} & 1
\end{pmatrix},
\nonumber\\
&& U_3 = u_0
\begin{pmatrix}
1 & \omega^{\xi} \\
\omega^{-\xi} & 1
\end{pmatrix},
\label{eq_submatrices}
\end{eqnarray}
%where $\Vec{K}_\xi^{(l)} = -\xi(2a_1^{*(l)}+a_2^{*(l)})/3$
where $\Vec{K}_\xi^{(l)}$
is the Dirac point of the layer $l$,
$\GVec{\sigma} = (\sigma_x, \sigma_y)$ is the Pauli matrices,
$u_0 \approx 0.11\, \rm eV$ represents
the in-plane Fourier component
of the interlayer transfer integral,
and $\omega = \rm exp(2 \pi i / 3)$.
For a moderate bias
such that
$\vare_\mathrm{D}^{(1)} \le -U/2 + \hbar v q$,
where
$|q|=|\Vec{K}_\xi^{(1)} - \Vec{K}_\xi^{(2)}|
=|\Vec{G}_1^{\rm M}|/\sqrt{3}
\approx 8 \pi \sin (\theta/2)/(3a)$,
the secular equation $\mathrm{det}(E-H_{\rm eff}(\Vec{k}))=0$
at $\Vec{k} = \Vec{K}_\xi^{(1)}$
is reduced to
\begin{equation}
E-\frac{U}{2} = \frac{6u_0^2}{(E+U/2)^2-(\hbar v q)^2} (E+\frac{U}{2}).
\label{eq_secular_det}
\end{equation}
Assuming also $6u_0^2 >> E^2, U^2/4$
for a moderate bias,
the Eq.\ (\ref{eq_secular_det})
finally gives an analytic expression Eq.\ (\ref{eq_analytic}).

\bibliography{Qiqqa2BibTexExport}

\end{document}